\documentclass[aps,prb,twocolumn,superscriptaddress]{revtex4}
\usepackage{graphicx}% Include figure files
\usepackage{epsfig}
\usepackage{dcolumn}% Align table columns on decimal point
\usepackage{bm}% Bold math
\usepackage{tabularx}% Bold math
\usepackage{hyperref}
\usepackage{xcolor}
\usepackage{amsmath}
\usepackage{subfigure}
\usepackage{float}
\usepackage{booktabs,dcolumn}
\usepackage{caption}
\usepackage{adjustbox}
\hypersetup{citecolor=black, filecolor= black, linkcolor= black, urlcolor= black}
\newcommand*{\rom}[1]{\expandafter\@slowromancap\romannumeral #1@}

\begin{document}

\title{Theoretical study of the pressure-induced structure, phase transition, mechanical and electronic properties in the V-N system}

\author{Jin Zhang}
\email{zhangjin225@gmail.com}
\affiliation{Department of Geosciences, Center for Materials by Design, and Institute for Advanced Computational Science, State University of New York, Stony Brook, NY 11794-2100, USA}

\author{Xinfeng Li}
\affiliation{Sino-French Institute of Nuclear Engineering and Technology, Sun Yat-sen University, Zhuhai, Guangzhou 519082, China}

\author{Xiao Dong}
\affiliation{Key Laboratory of Weak-Light Nonlinear Photonics and School of Physics, Nankai University, Tianjin 300071, China}

\author{Huafeng Dong}
\affiliation{School of Physics and Optoelectronic Engineering, Guangdong University of Technology, Guangzhou 510006, China}

\author{Artem R. Oganov}
\email{a.oganov@skoltech.ru}
\affiliation{Skolkovo Institute of Science and Technology, Skolkovo Innovation Center, 3 Nobel St., Moscow, 143026, Russia}
\affiliation{Moscow Institute of Physics and Technology, 9 Institutskiy Lane, Dolgoprudny 141700, Russia}
\affiliation{International Center for Materials Discovery, Northwestern Polytechnical University, Xi'an, Shaanxi 710072, PR China}

\begin{abstract}
Stable compounds in the V-N system are systematically searched and four new high-pressure phases are found, including \emph{C}2/\emph{m}-V$_9$N, \emph{Pbam}-V$_5$N$_2$, \emph{Pnma}-V$_2$N and \emph{I}4/\emph{mcm}-VN$_2$. V$_2$N undergoes a phase transition from $\varepsilon$-Fe$_2$N-type V$_2$N (\emph{P}$\bar{3}$1\emph{m}) to $\zeta$-Fe$_2$N-type V$_2$N (\emph{Pbcn}) at 10 GPa and to Fe$_2$C-type V$_2$N (\emph{Pnnm}) at 59 GPa, then to \emph{Pnma}-V$_2$N at 96 GPa. Low-temperature tetragonal VN is theoretically proved to belong to space group \emph{P}$\bar{4}$2\emph{m}. The estimated Vickers hardnesses and fracture toughness of WC-type VN are around 37 GPa and 4.3-6.1 MPa m$^{1/2}$, respectively. Al$_2$Cu-type VN$_2$ (\emph{I}4/\emph{mcm}) with a Vickers hardness of 25-27 GPa and fracture toughness of 3.6-6.6 MPa m$^{1/2}$ also shows excellent mechanical properties. Elastic properties of WC-type mononitrides of transition metals from IVB group (Ti, Zr and Hf), VB group (V, Nb and Ta) and VIB (Cr, Mo and W) are calculated and compared. Both the bond strength and structural configuration determine the mechanical properties of a material.
\end{abstract}
\maketitle

\section{Introduction}

Compared with pure metals, transition metal nitrides have extremely strong and short bonds which lead to a very low compressibility and high hardness. Therefore, a great deal of work has been performed to the search for new transition metal nitrides (and also carbides and borides) with extreme hardness and good fracture toughness, in the hope that many of these will find applications in cutting tools and as hard coatings. Among these metal nitrides, vanadium nitride becomes an attractive candidate because of its high hardness\cite{farges1993crystallographic,sanjines1998hexagonal}, high melting point\cite{rostoker1954survey}, good corrosion resistance\cite{escobar2014corrosion} and low friction coefficient\cite{wiklund2006evaporated,fallqvist2013influence}. Based on the combination of these outstanding physical and chemical properties, vanadium nitride thin films have been fabricated to be used as hard coatings\cite{kutschej2007experimental,caicedo2011mechanical} and nanocrystalline vanadium nitride can become an attractive material for supercapacitors\cite{choi2006fast}.

Properties are closely related to the structure of materials. For the crystal structures of V-N compounds, it has been established that $\beta$-V$_{2}$N$_{1-x}$ has a $\epsilon$-Fe$_{2}$N-type (\emph{P}$\bar{3}$1\emph{m}-V$_{2}$N) structure and $\delta$-VN$_{1-x}$ has a NaCl-type (\emph{Fm}$\bar{3}$\emph{m}-VN) \cite{carlson1986vanadium}. $\delta^{'}$-VN$_{1-x}$ (V$_{32}$N$_{26}$) was reported to exist below ~520 $^{\circ}$C and the lattice constant of its structure is twice as large as that of the original NaCl-type cell with six nitrogen atoms removed from the unit cell\cite{onozuka1978vacancy}. In addition, three metastable phases with stoichiometries V$_{16}$N, V$_{8}$N and V$_{9}$N$_{2}$ have been reported\cite{potter1974metastable,nouet1980domain} and can be formed according to a proposed metastable phase diagram \cite{carlson1986vanadium}. Other intermediate phases, including V$_{13}$N, V$_{9}$N and V$_{4}$N are believed to metastably exist\cite{nouet1980domain}.

Much effort has been made to explore the structures of these stable or metastable vanadium nitrides. Even for the simple vanadium mononitride, a significant amount of attention has been paid to study its crystal structure. Three decades ago, Kubel's\cite{kubel1988structural} experiment detected that VN crystallizes in the NaCl-type structure at 298 K and transforms into a low-temperature tetragonal-VN (a distorted NaCl-type phase, space group: \emph{P}$\bar{4}$2\emph{m}) at 205 K. Afterward, Weber's inelastic neutron scattering experiment\cite{weber1979phonon} found that NaCl-type VN exhibits a notably soft mode in its acoustic branch at room temperature. Meanwhile, first-principles calculations showed that NaCl-type VN is dynamically unstable according to its phonon dispersion curves\cite{isaev2007phonon,ivashchenko2008phonon} and also indicated that WC-type VN (\emph{P}$\bar{6}$\emph{m}2-VN) has the lowest energy at zero pressure\cite{ravi2009first,ivashchenko2011first,ravi2010cluster}. To explore the reason that NaCl-type VN is experimentally detected at room temperature while it is dynamically unstable from the first-principles calculation at 0 K. Ivashchenko's density functional theory (DFT) calculation\cite{ivashchenko2011first} proposed that vacancies stabilize NaCl-type VN compared to WC-type VN. Mei\cite{mei2015dynamic}, by performing both experiment and \emph{ab initio} molecular dynamics, found that tetragonal-VN appeared below 250 K and NaCl-type VN was thermodynamically stabilized by the temperature-induced anharmonic effects.

However, despite abundant theoretical and experimental studies of V-N compounds, less information is provided on the pressure-induced new compounds in the V-N system. Moreover, except that the hardness of bulk \emph{Fm}$\bar{3}$\emph{m}-VN (13 GPa) is reported by previous experiments\cite{toth2014transition,pierson1996handbook}, the mechanical properties of other V-N compounds are still experimentally lacking due to the difficulty of preparing the samples with different stoichiometries. In this work, we perform a comprehensive calculation to investigate the V-N system at pressure up to 120 GPa, analyze the dynamical stability of high-pressure phases, calculate the elastic constants of these high-pressure phases and then predict their hardness and fracture toughness at 0 GPa. Indeed, several new stoichiometries in the V-N system have been predicted under high pressure and they were found to possess unusual mechanical properties.

\section{Computational methodology}

Stable phases in the V-N system were searched using first-principles evolutionary algorithm (EA) as implemented in the USPEX code\cite{oganov2006crystal,lyakhov2013new,oganov2011evolutionary} combined with \emph{ab initio} structure relaxations using DFT with the Perdew--Burke--Ernzerhof (PBE) generalized gradient approximation (GGA) exchange--correlation functional\cite{perdew1996generalized}, as implemented in the VASP package\cite{kresse1996efficient}. We also tested the DFT+U method, but the calculated enthalpies of formation from pure DFT agree better with the experimental data at 0 K from NIST-JANAF thermochemical tables\cite{chase1998nist}. The variable-composition structure searches\cite{oganov2011evolutionary} were performed for the V-N system at 0 GPa, 10 GPa, 20 GPa, 30 GPa, 40 GPa, 50 GPa, 60 GPa, 70 GPa, 80 GPa, 90 GPa, 100 GPa, 110 GPa and 120 GPa. Several fixed-composition searches were also performed for vanadium mononitride at 0 GPa and for low-enthalpy metastable compounds found at various pressures. The initial generation of structures was produced randomly using space group symmetry, each subsequent generation was obtained by variation operators including heredity (40\%), lattice mutation (20\%), random generator (20\%) and transmutation (20\%). The electron-ion interaction was described by the projector-augmented wave (PAW) potentials\cite{blochl1994projector}, with 3\emph{p}$^6$4\emph{s}$^1$3\emph{d}$^4$ and 2\emph{s}$^2$2\emph{p}$^3$ shells treated as valence for V and N, respectively. The plane-wave energy cutoff was chosen as 600 eV and $\Gamma$-centered uniform \emph{k}-meshes with resolution 2$\pi\times0.06$ {\AA}$^{-1}$ were used to sample the Brillouin zone. Phonon dispersions were calculated using the finite-displacement method implemented in the Phonopy code\cite{togo2008first}, based on the Hellmann-Feynman forces calculated with the VASP code. Voight-Reuss-Hill approximation has been adopted to estimate the polycrystal bulk modulus (\emph{B}), and shear modulus (\emph{G}). The Vickers hardness was estimated according to the Chen-Niu's hardness model\cite{chen2011modeling} and its modification by Tian\cite{tian2012microscopic}, as well as Mazhnik's hardness model\cite{mazhnik2019model}. The fracture toughness was evaluated by Niu's fracture toughness model of covalent and ionic crystals\cite{niu2019simple} and Mazhnik's fracture toughness model\cite{mazhnik2019model}. The calculation of crystal orbital Hamilton population (COHP) was performed using the TB-LMTO-ASA program\cite{krier1995tb} based on the binding linear muffin-tin orbital method with the atomic sphere approximation.

\section{Results and discussions}
\subsection{Crystal structure prediction for V-N system}

\begin{center}
\begin{figure*}
   \includegraphics[angle=0,width=0.9\linewidth]{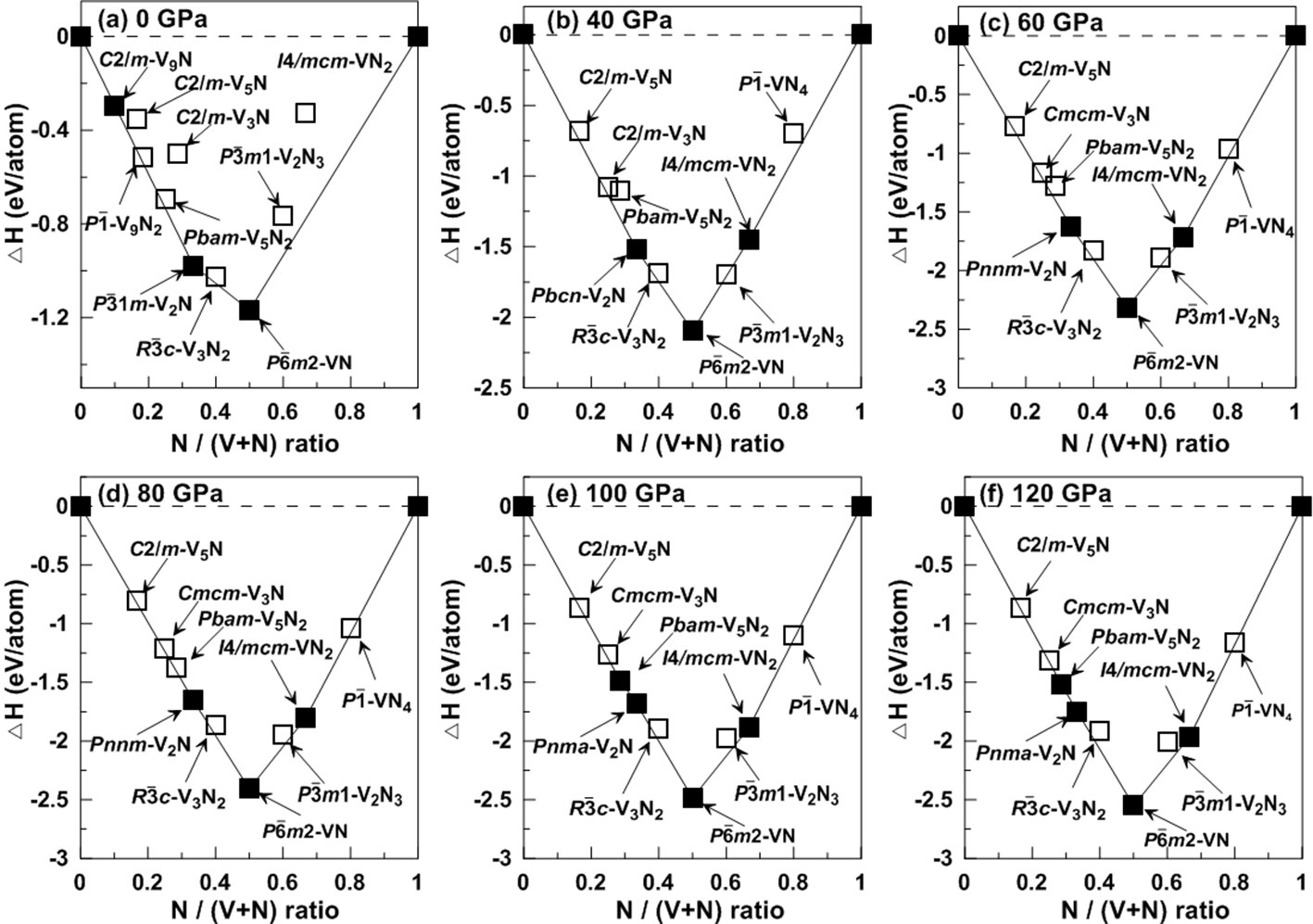}
   \caption{\label{Fig0} Convex-hull diagrams for the V--N system at (a) 0 GPa, (b) 40 GPa, (c) 60 GPa, (d) 80 GPa, (e) 100 GPa, and (f) 120 GPa, respectively. Solid squares denote stable phases while open squares represent metastable compounds.}
\end{figure*}
\end{center}

\begin{center}
\begin{figure}[H]
   \includegraphics[angle=0,width=0.9\linewidth]{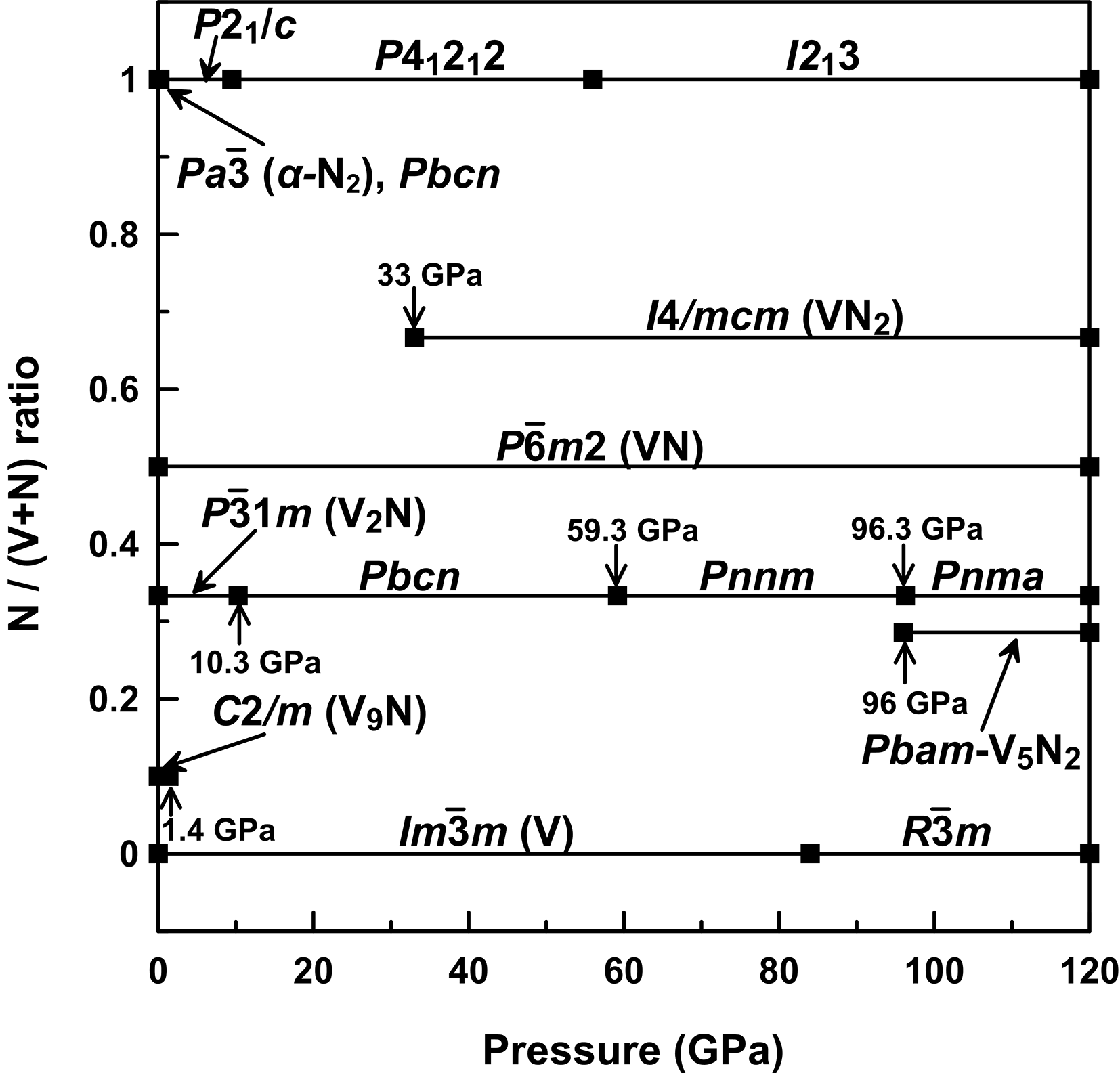}
   \caption{\label{Fig1} Pressure-composition phase diagram of the V-N system at pressures up to 120 GPa.}
\end{figure}
\end{center}

\begin{center}
\begin{figure}
   \includegraphics[angle=0,width=1.0\linewidth]{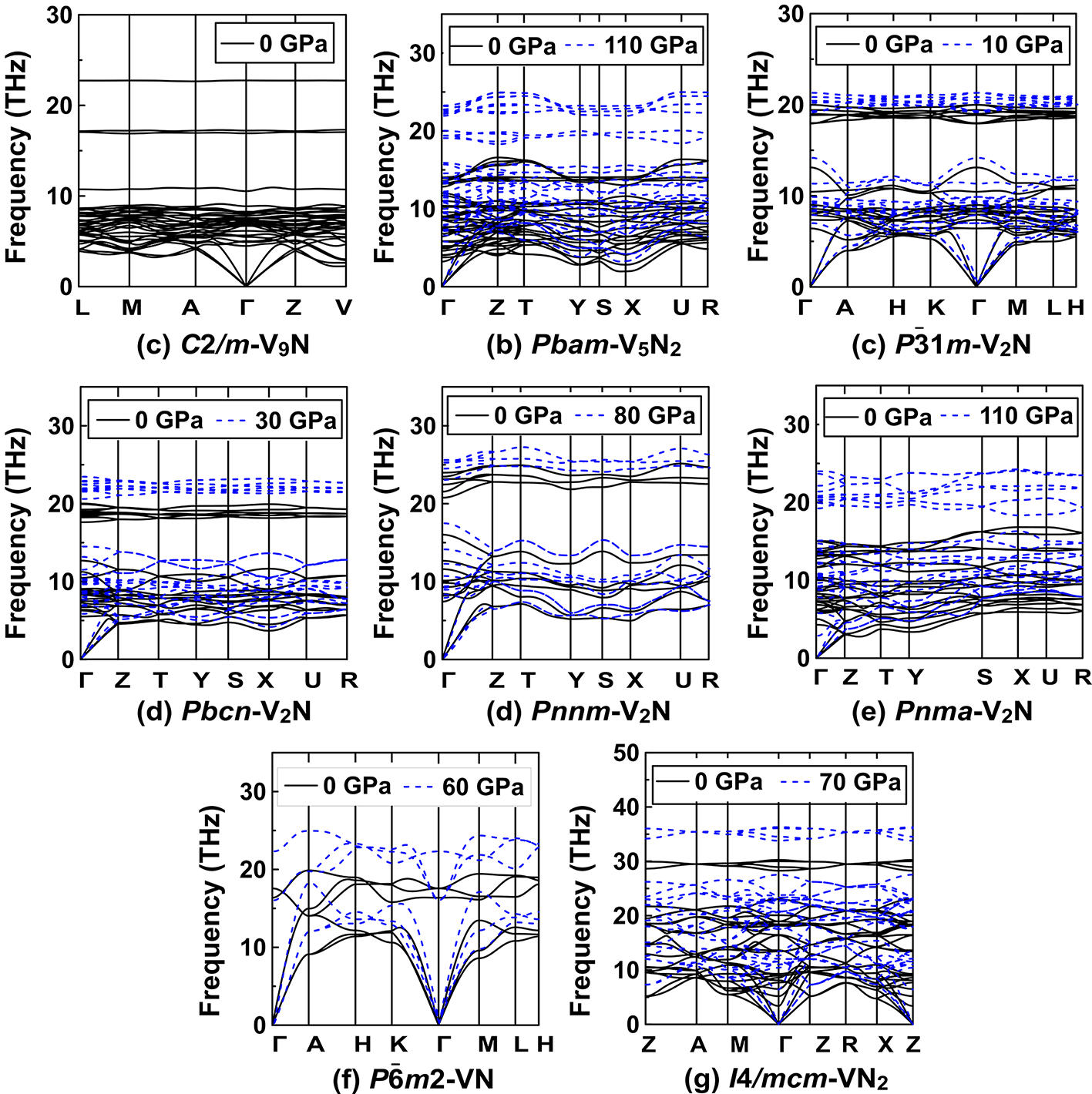}
   \caption{\label{Fig2} (Color online) Calculated phonon dispersion curves of V-N compounds along high-symmetry directions of the Brillouin zone. The solid black and dashed blue lines represent the results at zero and high pressures, respectively.}
\end{figure}
\end{center}

The pressure-composition phase diagram of the V-N system, shown in Fig. \ref{Fig1}, was constructed according to the calculated convex-hull diagrams [Fig.\ref{Fig0}]. Three compounds, \emph{C}2/\emph{m}-V$_9$N, \emph{P}$\bar{3}$1\emph{m}-V$_2$N and \emph{P}$\bar{6}$\emph{m}2-VN, are stable at zero pressure. Three new compounds are found under high pressure: \emph{I}4/\emph{mcm}-VN$_2$ (Al$_2$Cu-type) appears at 33 GPa and is stable at least up to 120 GPa; \emph{Pbam}-V$_5$N$_2$ becomes stable at 84 GPa; \emph{Pnma}-V$_2$N is a new high-pressure phase of V$_2$N. The dynamical stabilities of all the V-N compounds were checked by calculating phonon dispersions [Fig. \ref{Fig2}], no imaginary vibrational frequencies are found in the whole Brillouin zone at 0 GPa, indicating all these high-pressure vanadium nitrides can be theoretically preserved as metastable phases at ambient pressure.

\begin{center}
\begin{figure}
   \includegraphics[angle=0,width=0.9\linewidth]{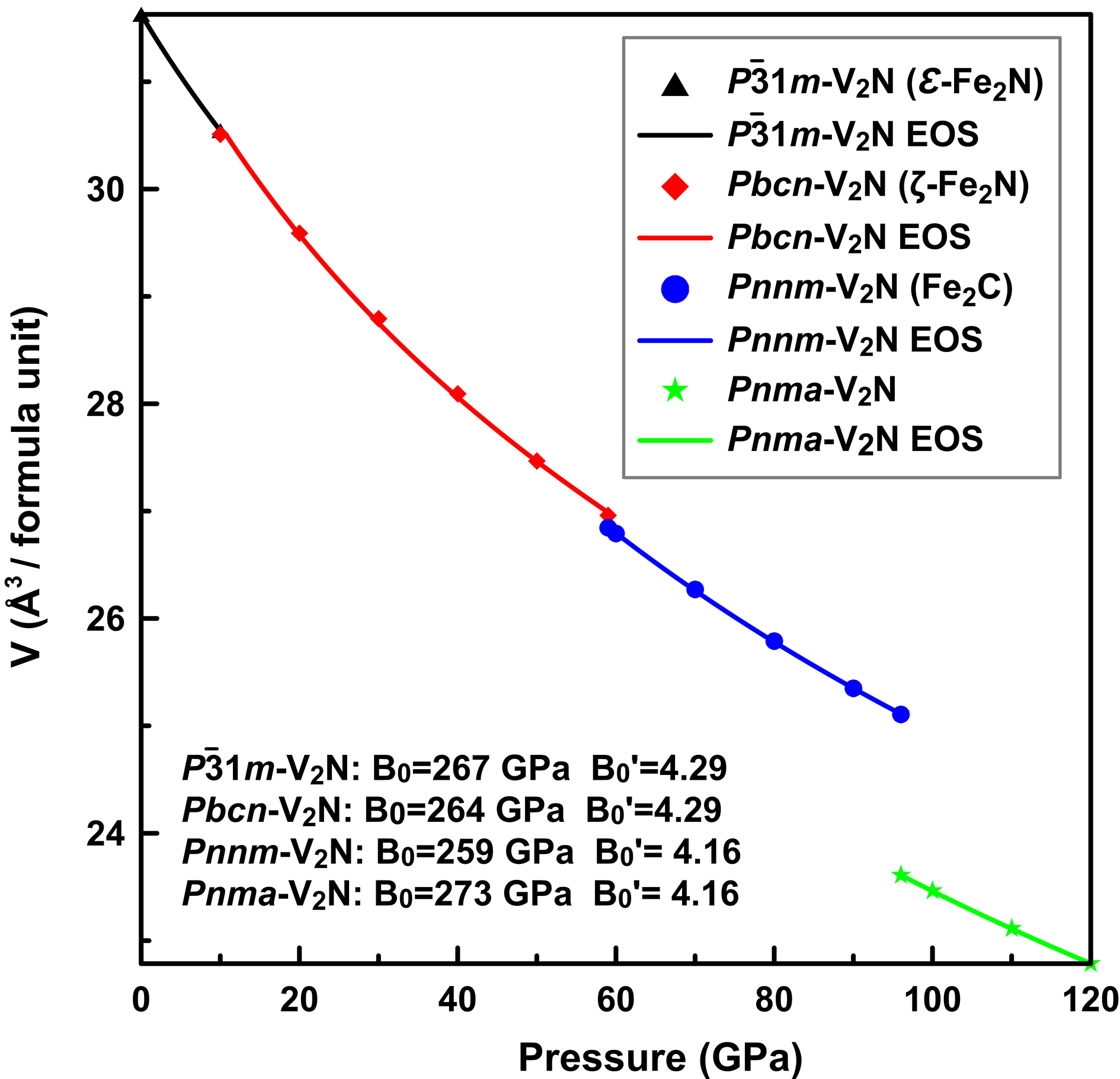}
   \caption{\label{Fig3} (Color online) Equation of state (EOS) of V$_{2}$N up to 120 GPa. The calculated pressure-volume data were fit with a third-order Birch-Murnaghan equation of state \emph{P}(V)=(3\emph{B}$_{0}/2$)[(\emph{V}/\emph{V}$_{0}$)$^{-7/3}$-(\emph{V}/\emph{V}$_{0}$)$^{-5/3}$]\{1+(3/4)(\emph{B}$_{0}^{'}$-4)[(\emph{V}/\emph{V}$_{0}$)$^{-2/3}$-1]\} to find \emph{B}$_{0}$ and \emph{B}$_{0}^{'}$.}
\end{figure}
\end{center}

In 1949, $\varepsilon$-Fe$_2$N-type (\emph{P}$\bar{3}$1\emph{m}) V$_2$N structure was proposed by Hahn\cite{hahn1949system} and verified by neutron diffraction of melting vanadium nitride in nitrogen at 1 MPa\cite{christensen1979structure}. Recently, Ravi\cite{ravi2009first} performed DFT calculation to study the property and phase stability of $\zeta$-Fe$_2$N-type V$_2$N (\emph{Pbcn}), Fe$_2$C-type V$_2$N (\emph{Pnnm}) and CdI$_2$-type V$_2$N (\emph{P}$\bar{3}$\emph{m}1). Our calculated results show that both $\zeta$-Fe$_2$N-type and Fe$_2$C-type V$_2$N are actually the high-pressure phases of V$_2$N and V$_2$N undergoes a series of structural transformation with increasing pressure: $\varepsilon$-Fe$_2$N-type V$_2$N first transforms into the $\zeta$-Fe$_2$N-type V$_2$N at 10 GPa, then to Fe$_2$C-type-V$_2$N at 59 GPa, and then to the \emph{Pnma}-V$_2$N at 96 GPa, as displayed in Fig. \ref{Fig3}. In the $\varepsilon$-Fe$_2$N-type V$_2$N, $\zeta$-Fe$_2$N-type V$_2$N and Fe$_2$C-type V$_2$N structures, all the V atoms are three-coordinate and N atoms are six-coordinate while the coordination numbers of V and N atoms in \emph{Pnma}-V$_2$N increase to four and eight, respectively.

Previous experimental studies reported that the NaCl-type VN transforms into a tetragonal VN on cooling and the transition temperature is 204 K for bulk polycrystalline VN samples\cite{kubel1988structural} and 250-300 K for VN films\cite{mei2015dynamic}. In our calculations, besides the known WC-type VN (\emph{P}$\bar{6}$\emph{m}2), NiAs-type VN (\emph{P}6$_3$/\emph{mmc}), NaCl-type VN (\emph{Fm}$\bar{3}$\emph{m}) and CsCl-type VN (\emph{Pm}$\bar{3}$\emph{m}), we also successfully found the \emph{P}$\bar{4}$2\emph{m}-VN which was proposed by Kubel\cite{kubel1988structural}; the \emph{P}4$_2$\emph{mc}-VN and \emph{I}4$_1$\emph{md}-VN which were reported by Pu's first-principles calculation\cite{chun2014elastic}; the \emph{P}4$_2$/\emph{mcm}-VN, and \emph{P}4/\emph{nmm}-VN that were predicted in Ivashchenko's first-principles study\cite{ivashchenko2008phonon}. Furthermore, new low-energy phases: \emph{R}3\emph{m}-VN, \emph{Imm}2-VN and \emph{Cmc}2$_1$-VN were also found. Except NaCl-type VN, which is dynamically unstable at zero pressure and temperature based on the theoretical analysis of its lattice dynamics\cite{isaev2007phonon}, the dynamical stabilities of all the other vanadium mononitride were checked by calculating phonon dispersions (see Fig. S1 in the Supplementary Materials) at 0 GPa. The results show that the imaginary frequencies are found in phonon dispersion curves of \emph{R}3\emph{m}-VN, \emph{P}4/\emph{nmm}-VN and CsCl-type VN, indicating these three structures are dynamically unstable. Then the zero-point energies (ZPE) were calculated for the dynamically stable structures (see relative enthalpy$+$ZPE values in Table \ref{Tab1}). Total enthalpy sequence (including ZPE) for the eight dynamically stable sructures is WC-type VN $<$ \emph{P}4$_2$\emph{mc}-VN $<$ \emph{I}4$_1$\emph{md}-VN $<$ AsNi-type VN $<$ \emph{Imm}2-VN $<$ \emph{Cmc}2$_1$-VN $<$ \emph{P}$\bar{4}$2\emph{m}-VN $<$ \emph{P}4$_2$/\emph{mcm}-VN. Among them, \emph{P}4$_2$\emph{mc}-VN, \emph{I}4$_1$\emph{md}-VN, \emph{P}$\bar{4}$2\emph{m}-VN and \emph{P}4$_2$/\emph{mcm}-VN all belong to tetragonal system. From the viewpoint of energetic stability, it seems that \emph{P}4$_2$\emph{mc}-VN is more promising than experimentally proposed \emph{P}$\bar{4}$2\emph{m}-VN to be that tetragonal-VN. Is this really the case?

\begin{center}
\begin{figure}
   \includegraphics[angle=0,width=0.9\linewidth]{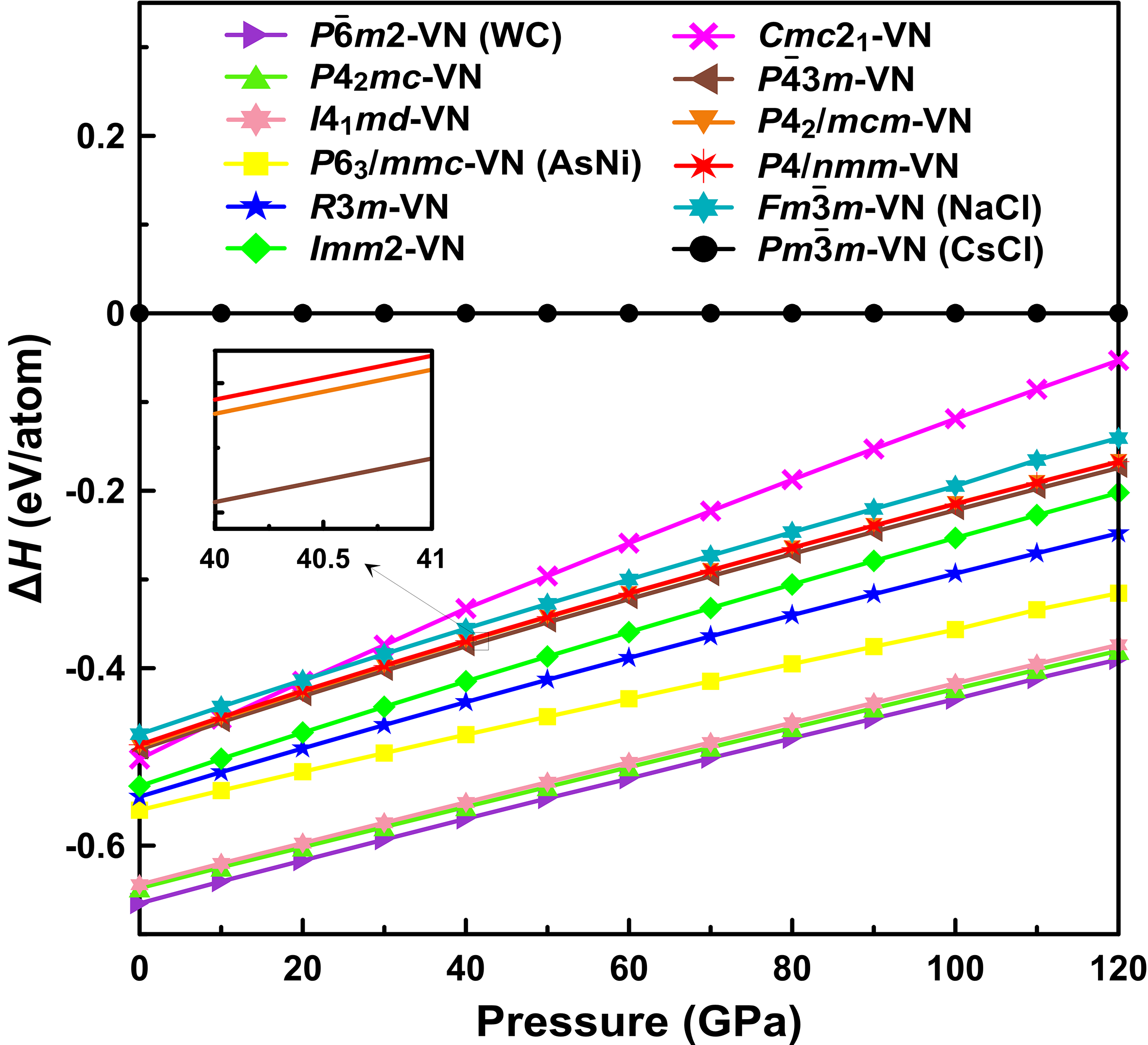}
   \caption{\label{Fig5} (Color online) Relative enthalpy as a function of pressure for vanadium mononitrides. CsCl-type VN is taken as reference.}
\end{figure}
\end{center}

\begin{table}
\centering
\caption{Relative enthalpy (taking the enthalpy of \emph{P}$\bar{6}$\emph{m}2-VN as reference) without and with zero-point energy (ZPE), dynamical stability (DS) for low-energy vanadium mononitrides at 0 GPa. ZPE can only be calculated for the dynamically stable phases. $\Delta$E denotes the relative enthalpy without ZPE and $\Delta$E$^{*}$ represents the relative enthalpy with ZPE. ${\surd}$ = dynamical stability; ${\times}$ = dynamical instability; NA = not available.}
\begin{tabular}{ c c c  c }   % Alignment for each cell: l=left, c=center, r=right
\hline \hline
Compound                        &$\Delta$E  & DS        & $\Delta$E$^{*}$  \\
                                &(eV/atom)  &           & (eV/atom)    \\
\hline
\emph{P}$\bar{6}$\emph{m}2-VN   &   0       & ${\surd}$  &   0   \\
\emph{P}4$_2$\emph{mc}-VN       & 0.01843  & ${\surd}$  & 0.00097   \\
\emph{I}4$_1$\emph{md}-VN       & 0.02239  & ${\surd}$  & 0.00518    \\
\emph{P}6$_3$/\emph{mmc}-VN     & 0.10445  & ${\surd}$  & 0.08094    \\
\emph{R}3\emph{m}-VN            & 0.12156  & ${\times}$ & NA          \\
\emph{Imm}2-VN                  & 0.13347  & ${\surd}$  & 0.10848    \\
\emph{Cmc}2$_1$-VN              & 0.16409 & ${\surd}$   & 0.13914    \\
\emph{P}$\bar{4}$2\emph{m}-VN   & 0.17469  & ${\surd}$  & 0.13714  \\
\emph{P}4$_2$/\emph{mcm}-VN     & 0.17505  & ${\surd}$  & 0.13846   \\
\emph{P}4/\emph{nmm}-VN         & 0.18003  & ${\times}$ & NA           \\
\emph{Fm}$\bar{3}$\emph{m}-VN   & 0.19205  & ${\times}$ & NA           \\
\emph{Pm}$\bar{3}$\emph{m}-VN   & 0.67864  & ${\times}$ & NA           \\
\hline\hline
\end{tabular}
\label{Tab1}
\end{table}

\begin{center}
\begin{figure*}
   \includegraphics[angle=0,width=0.9\linewidth]{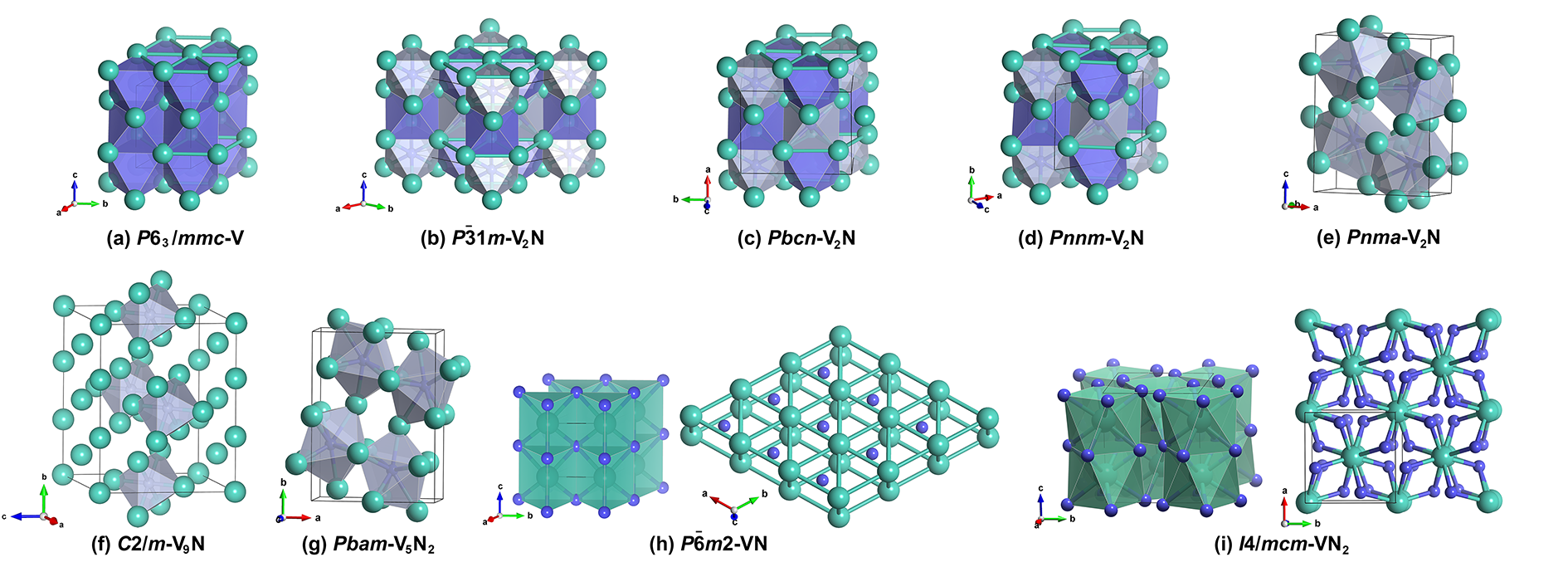}
   \caption{\label{Fig6} (Color online) Crystal structures of (a) \emph{P}6$_3$/\emph{mmc}-V; (b) \emph{P}$\bar{3}$1\emph{m}-V$_2$N; (c) \emph{Pbcn}-V$_2$N; (d) \emph{Pnnm}-V$_2$N; (e) \emph{Pnma}-V$_2$N; (f) \emph{C}2/\emph{m}-V$_9$N; (g) \emph{Pbam}-V$_5$N$_2$; (h) \emph{P}$\bar{6}$\emph{m}2-VN ; (i) \emph{I}4/\emph{mcm}-VN$_2$. The hcp sublattices of vanadium are highlighted in polyhedral representation of (a), (b), (c) and (d). Nitrogen-centered octahedra and nitrogen vacancies are shown as light and dark purple polyhedra, respectively. Vanadium-centered polyhedra are shown in green.}
\end{figure*}
\end{center}

To make this question clear, the selected-area electron diffraction patterns of tetragonal vanadium mononitrides (\emph{P}4$_2$\emph{mc}-VN, \emph{I}4$_1$\emph{md}-VN, \emph{P}$\bar{4}$2\emph{m}-VN, \emph{P}4$_2$/\emph{mcm}-VN and \emph{P}4/\emph{nmm}-VN) were calculated using CrystalMaker software\cite{palmer2007crystalmaker}. It turns out that only the calculated selected-area electron diffraction patterns of \emph{P}$\bar{4}$2\emph{m}-VN perfectly match the experimental electron diffraction of tetragonal-VN\cite{mei2015dynamic}, see the Fig. S4 in Supplementary Materials. Therefore, we can confirm that the experimentally observed tetragonal-VN is \emph{P}$\bar{4}$2\emph{m}-VN. Nevertheless, why the tetragonal-VN is identical to \emph{P}$\bar{4}$2\emph{m}-VN, and not to the more energetically favorable tetragonal \emph{P}4$_2$\emph{mc}-VN or \emph{I}4$_1$\emph{md}-VN? The most probable reason is the strong anharmonicity of VN\cite{mei2015dynamic}. The enthalpy-pressure diagram (without ZPE) of vanadium mononitrides is drawn in Fig. \ref{Fig5}. WC-type VN structure, which was never reported by the experiment, might be synthesized through the application of pressure.

\subsection{Structures of V-N compounds}

The optimized structural parameters for V-N compounds are listed in Table S1 in Supplementary Materials. \emph{C}2/\emph{m}-V$_9$N and \emph{Pbam}-V$_5$N$_2$, which respectively have six-coordinate N and eight-coordinate N in their structures, cannot be derived from close packing. The hexagonal structure of WC-type VN has V atoms in a sixfold trigonal prismatic coordination, as displayed in Fig. \ref{Fig6}(h).

\emph{P}$\bar{3}$1\emph{m}-V$_2$N, \emph{Pbcn}-V$_2$N and \emph{Pnnm}-V$_2$N are formed by nitrogen atoms occupying the octahedral interstitial sites of hcp vanadium sublattice. The polyhedral representations of these V$_2$N structures are shown in Fig. \ref{Fig6}(b) (c) (d) and hcp vanadium [Fig. \ref{Fig6}(a)] is drawn for comparison. The structure of \emph{P}$\bar{3}$1\emph{m}-V$_2$N also derives from hcp vanadium, in which nitrogen atoms occupy the octahedral interstitial sites. Close-packed vanadium sublattice is no longer observed in the higher-pressure phase \emph{Pnma}-V$_2$N [see Fig. \ref{Fig6}(e)].

The crystal structure of high-pressure VN$_2$ phase is similar with that of high-pressure TiN$_2$\cite{yu2015phase}. Both of them have the Al$_2$Cu-type (\emph{I}4/\emph{mcm}) structure. This is different with the case in transition metal borides where the Al$_2$Cu-type structure is often formed with a transition metal /boron ratio of 2:1 (e.g. Ti$_2$B, Cr$_2$B, Mn$_2$B, Fe$_2$B, Co$_2$B, Ni$_2$B and Mo$_2$B)\cite{mohn1988calculated} while the AlB$_2$-type (\emph{P}6/\emph{mmm}) structure is predominant for transition metal borides with a transition metal/boron ratio of 1:2 (e.g. TiB$_2$, VB$_2$, CrB$_2$, MnB$_2$, FeB$_2$, MoB$_2$, IrB$_2$)\cite{friedrich2011synthesis,vajeeston2001electronic}.

\subsection{Mechanical properties of V-N compounds}

The elastic moduli, including bulk modulus $\emph{B}$, shear modulus $\emph{G}$ and Young's modulus $\emph{E}$ of V-N compounds are calculated from the elastic tensors, and theoretical hardnesses and fracture toughness [listed in Table \ref{Tab2}] can be further predicted based on the empirical models\cite{chen2011modeling,tian2012microscopic,niu2019simple,mazhnik2019model}. All the V-N compounds have high bulk moduli. The mechanical stabilities of the vanadium nitrides have been checked by calculating their elastic constants and all these compounds are mechanically stable at 0 GPa based on the Born criteria of mechanical stability\cite{nye1985physical}.

Fracture toughness describes the resistance of a material against crack propagation and is one of the most critical mechanical properties of materials. Design of a material with excellent mechanical properties requires both high hardness and high fracture toughness. WC-type VN possess the best combination of hardness ($\sim$37 GPa) and fracture toughness (4.3-6.1 MPa m$^{1/2}$). The second hardest V-N compound is \emph{I}4/\emph{mcm}-VN$_2$ which also has good fracture toughness with the value of 3.6-6.6 MPa m$^{1/2}$. \emph{C}2/\emph{m}-V$_9$N exhibits poor fracture toughness and its hardness is around 3.5-9.8 GPa.

\begin{table*}
\begin{center}
\captionsetup{font=footnotesize}{
\caption{\label{Tab2} Calculated bulk modulus $\emph{B}$, shear modulus $\emph{G}$, Young's modulus $\emph{E}$, Poisson's ratio $\upsilon$, volume per atom \emph{V}$_{0}$, Vickers hardness \emph{H}$_{v}$ (\emph{H}$_{v}^{c}$ is calculated from Chen-Niu's model, \emph{H}$_{v}^{t}$ is calculated from Tian's modification model\cite{tian2012microscopic}, \emph{H}$_{v}^{m}$ is calculated from Mazhnik's model\cite{mazhnik2019model}) and fracture toughness \emph{K}$_{IC}$ (\emph{K}$_{IC}^{n}$ is calculated from Niu's model\cite{niu2019simple} , \emph{K}$_{IC}^{m}$ is calculated from Mazhnik's model\cite{mazhnik2019model}) of V-N compounds at 0 GPa, as well as literature values for \emph{Fm}$\bar{3}$\emph{m}-VN. \emph{G}/\emph{B} and $\upsilon$ are dimensionless; \emph{K}$_{IC}$ is in MPa m$^{1/2}$; \emph{V}$_{0}$ is in ${\AA}{}^{3}/atom$. Hardness, $\emph{B}$, $\emph{G}$ and $\emph{E}$ are in GPa.}}

\resizebox{\textwidth}{!}{
\begin{tabular}{ c c c  c c c  c c c  c c c  c c c  c c c c  c c c  c c   c}   % Alignment for each cell: l=left, c=center, r=right
\hline \hline
Compound &\emph{C}$_{11}$& \emph{C}$_{22}$&\emph{C}$_{33}$&\emph{C}$_{44}$&\emph{C}$_{55}$&\emph{C}$_{66}$&\emph{C}$_{12}$&\emph{C}$_{13}$&\emph{C}$_{14}$& \emph{C}$_{16}$&\emph{C}$_{23}$& \emph{C}$_{26}$& \emph{C}$_{36}$& \emph{B}$_H$  & \emph{G}$_H$ & \emph{E}  &  \emph{G}/\emph{B}  & $\upsilon$ &\emph{V}$_{0}$  & \emph{H}$_{v}^{c}$ &\emph{H}$_{v}^{t}$ &\emph{H}$_{v}^{m}$ & \emph{K}$_{IC}^{n}$  &\emph{K}$_{IC}^{m}$ \\
\hline
\emph{C}2/\emph{m}-V$_9$N         &278&314&282&49 &55 &87 &130&166&  & -7  &   &-3 &-1&193&65 &176&0.34&0.35& 12.72 &3.5 &5.2   & 9.8   &1.7  & 2.8 \\
\emph{Pbam}-V$_5$N$_2$            &351&400&417&153&117&93 &197&169&  &     &189&   &  &252&111&291 &0.44&0.31& 10.25 &9.1 &10.2 & 15.4  &2.5  & 4.3 \\
\emph{P}$\bar{3}$1\emph{m}-V$_2$N &453&   &441&149&   &   &162&182&26& &   & & &267&142&361&0.53&0.27& 10.54  &14.3&14.9         & 17.7  &2.9  & 4.7 \\
\emph{Pbcn}-V$_2$N                &424&465&459&147&158&155&180&176&  & &157& & &264&147&372&0.56&0.26& 10.53   &15.7&16.2         & 17.8  &2.9  & 5.0  \\
\emph{Pnnm}-V$_2$N                &473&419&472&145&148&180&171&147&  & &157& & &257&153&383&0.60&0.25& 10.52 &17.7&18.0         & 17.8  &2.9  & 4.4\\
\emph{Pnma}-V$_2$N                &385&450&457&153&129&130&208&188&  & &160& & &267&130&335&0.49&0.29& 9.82  &11.9&12.7         & 17.1  &2.7  & 4.7 \\
\emph{P}$\bar{6}$\emph{m}2-VN     &634&   &890&271&   &   &172&127&  & &   & & &332&267&631&0.80&0.18& 8.65  &37.6&37.4         & 37.7  &4.3  & 6.1  \\
\emph{Fm}$\bar{3}$\emph{m}-VN     &621&   &   &119&   &   &162&   &  & &   & & &315&156&401&0.49&0.29& 8.77  &13.8&14.7         & 20.3  &3.2  & 5.6  \\
                                  &616$^a$&   &   &165$^a$&   &   &163$^a$&   &  &  & & & &312$^a$&156$^a$&402$^a$&    &0.29$^a$& & 10-13$^b,$$^c$ &  &\\
\emph{P}$\bar{4}$2\emph{m}-VN     &510&   &510&127&   &127&198&198&  & &   & & &302&138&360&0.46&0.30& 8.77 &11.3&12.4         & 18.8  &2.9  & 5.1\\
%\emph{P}4$_2$\emph{mc}-VN         &774&   &629&240&   &285&129&145&  & &   & &334&268&635&0.80&0.18&37.8&37.6&4.3\\
%\emph{I}4$_1$\emph{md}-VN         &779&   &650&242&   &287&132&136&  & &   & &334&272&643&0.82&0.18&38.9&38.7&4.3\\
\emph{I}4/\emph{mcm}-VN$_2$       &631&   &609&168&   &292&212&127&  & & &   & &311&214&522&0.69&0.22& 7.71 &26.8&26.9         & 24.8  &3.6 & 6.6\\
\hline\hline
\end{tabular}
}
\end{center}
\footnotetext{$^a$Calculated GGA result\cite{liang2010effect}, $^b$Experiment\cite{toth2014transition},$^c$Experiment\cite{wang2016synthesis}.}
\end{table*}

Poisson's ratio $\upsilon$ correlates with the degree of metallicity: the higher value of $\upsilon$, the weaker covalent bonding in the structure, the more metallic the material, and the lower hardness of this material\cite{frantsevich1983elastic}. Superhard phases generally have a low Poisson's ratio (about 0-0.2)\cite{brazhkin2002harder}. This is in accordance with our calculated Poisson's ratio for V-N compounds (Table \ref{Tab2}), e.g. WC-type VN and \emph{I}4/\emph{mcm}-VN$_2$, which exhibit higher hardness, have low Poisson's ratios 0.18 and 0.22, respectively. Large values of \emph{C}$_{33}$ (890 GPa) and \emph{C}$_{44}$ (271 GPa) for WC-type VN indicate its extremely high incompressibility along the c-axis and great resistance to shear deformation. Pugh proposed a modulus ratio (\emph{G}/\emph{B} ratio) to discriminate between brittle and ductile phases\cite{pugh1954xcii}: a material is deemed to be brittle if \emph{G}/\emph{B} $>$ 0.57 and \emph{G}/\emph{B} $<$ 0.57 indicates a ductile material. Hence, WC-type VN, \emph{I}4/\emph{mcm}-VN$_2$ and \emph{Pnnm}-V$_2$N are brittle while \emph{Pbam}-V$_5$N$_2$, \emph{P}$\bar{3}$1\emph{m}-V$_2$N, \emph{Pbcn}-V$_2$N and \emph{Pnma}-V$_2$N are ductile.

\begin{table*}
\centering
\captionsetup{font=footnotesize}{
\caption{\label{Tab3} Calculated bulk modulus $\emph{B}$, shear modulus $\emph{G}$ and and hardness of WC-type structures for nitrogen combined with transitional elements from IVB (Ti, Zr and Hf), VB (V, Nb and Ta) and VIB (Cr, Mo and W) groups at 0 GPa. Except dimensionless \emph{G}/\emph{B} and \emph{K}$_{IC}$ in MPa m$^{1/2}$, all properties are in GPa. Superscript $*$ represents the structure is the metastable phase while $+$ represents the structure is the stable structure at zero pressure and temperature at the GGA level. \emph{H}$_{v}^{c}$ is calculated from Chen-Niu's model, \emph{H}$_{v}^{t}$ is calculated from Tian's modification model\cite{tian2012microscopic}, \emph{H}$_{v}^{m}$ is calculated from Mazhnik's model\cite{mazhnik2019model}, \emph{K}$_{IC}^{n}$ is calculated from Niu's model\cite{niu2019simple} and \emph{K}$_{IC}^{m}$ is calculated from Mazhnik's model\cite{mazhnik2019model}. Note the most stable structure for NbN is anti-WC-type structure instead of WC-type structure\cite{ivashchenko2010phase,zhao2015phase}, thus the superscript $*$ is used for WC-type NbN in this table.}}

\resizebox{\textwidth}{!}{
\begin{tabular}{ c c c c  c c c c  c c c c  c c c  c c c  c c c  c c c c c c}   % Alignment for each cell: l=left, c=center, r=right
\hline \hline
Compound & \emph{B}$_H$ & \emph{G}$_H$& \emph{G}/\emph{B}& \emph{H}$_{v}^{c}$ & \emph{H}$_{v}^{t}$ & \emph{H}$_{v}^{m}$ & \emph{K}$_{IC}^{n}$ & \emph{K}$_{IC}^{m}$ & Compound &\emph{B}$_H$ & \emph{G}$_H$ & \emph{G}/\emph{B} & \emph{H}$_{v}^{c}$ & \emph{H}$_{v}^{t}$ &  \emph{H}$_{v}^{m}$ & \emph{K}$_{IC}^{n}$ & \emph{K}$_{IC}^{m}$  &Compound &  \emph{B}$_H$ & \emph{G}$_H$ & \emph{G}/\emph{B} & \emph{H}$_{v}^{c}$ & \emph{H}$_{v}^{t}$ & \emph{H}$_{v}^{m}$  & \emph{K}$_{IC}^{n}$& \emph{K}$_{IC}^{m}$ \\
\hline
TiN$^{*}$ & 257 & 148 & 0.58& 16.5 &16.9& 17.6  &2.9& 3.6 & VN$^{+}$  & 332    & 267    & 0.8 & 37.6 &37.4& 37.7 &4.3& 6.1 &CrN$^{+}$  &  354  &  227  & 0.64 & 25.4&25.9 & 25.8 &4.0& 5.8\\
          &     &     &     &      &    &   &   &  &           &318$^a$ & 213$^a$& 0.67$^a$& &    &     &   & &    &  390$^f$&  348$^f$  & 0.89$^f$    &     &  &  & &\\
ZrN$^{*}$ & 222 & 116 & 0.52& 12.0 &12.7& 14.7  &2.5& 2.9 &NbN$^{*}$ & 312 & 217  & 0.70& 27.5 &27.6& 25.4  &3.9& 5.3  &MoN$^{*}$  &  347  &  153  & 0.44 & 11.5&12.8 & 21.2 &3.4& 5.0\\
          & 225$^d$& 118$^d$& 0.52$^d$& & &   & &&      & 316$^d$ & 209$^d$  & 0.66$^d$& &&   &   &&   &  351$^d$  & 181$^d$   & 0.52$^d$ &  &   && &\\
          &     &     &     &      &    &   &        &     &      &     &      &    &    &        &  345$^e$  &    &      &   &   &\\
HfN$^{*}$ & 240 & 139 & 0.58& 15.9 &16.2& 16.5  &2.8& 3.3 & TaN$^{+}$ & 337 & 247  & 0.73& 32.0 &32.0& 30.5 &4.3& 6.1  &WN$^{+}$   &  374  &  144  & 0.39 & 9.0 &10.5 & 21.0 &3.4& 5.3  \\
          & 253$^c$ & 137$^c$ & 0.54$^c$&&   &  & &&   & 384$^c$ & 261$^c$& 0.68$^c$& &  &&    &     & & 349$^c$ &148$^c$  & 0.42$^c$ &   &  & & &\\
          &         &           &       &&   &  & &&   & 318$^a$ & 213$^a$& 0.67$^a$& &  &&    &     & & 376$^b$ &157$^b$  &  0.42$^b$    &  &&   &    &\\
\hline\hline
\end{tabular}
}
\footnotetext{$^a$Calculation\cite{li2011crystal}, $^b$Calculation\cite{song2010first}, $^c$Calculation\cite{zhao2008electronic}, $^d$Calculation\cite{zhao2010structural}, $^e$Experiment\cite{ganin2006synthesis}, $^f$Calculation\cite{ming2015structural}.}
\end{table*}

\begin{center}
\begin{figure}[htbp]
   \includegraphics[angle=0,width=1\linewidth]{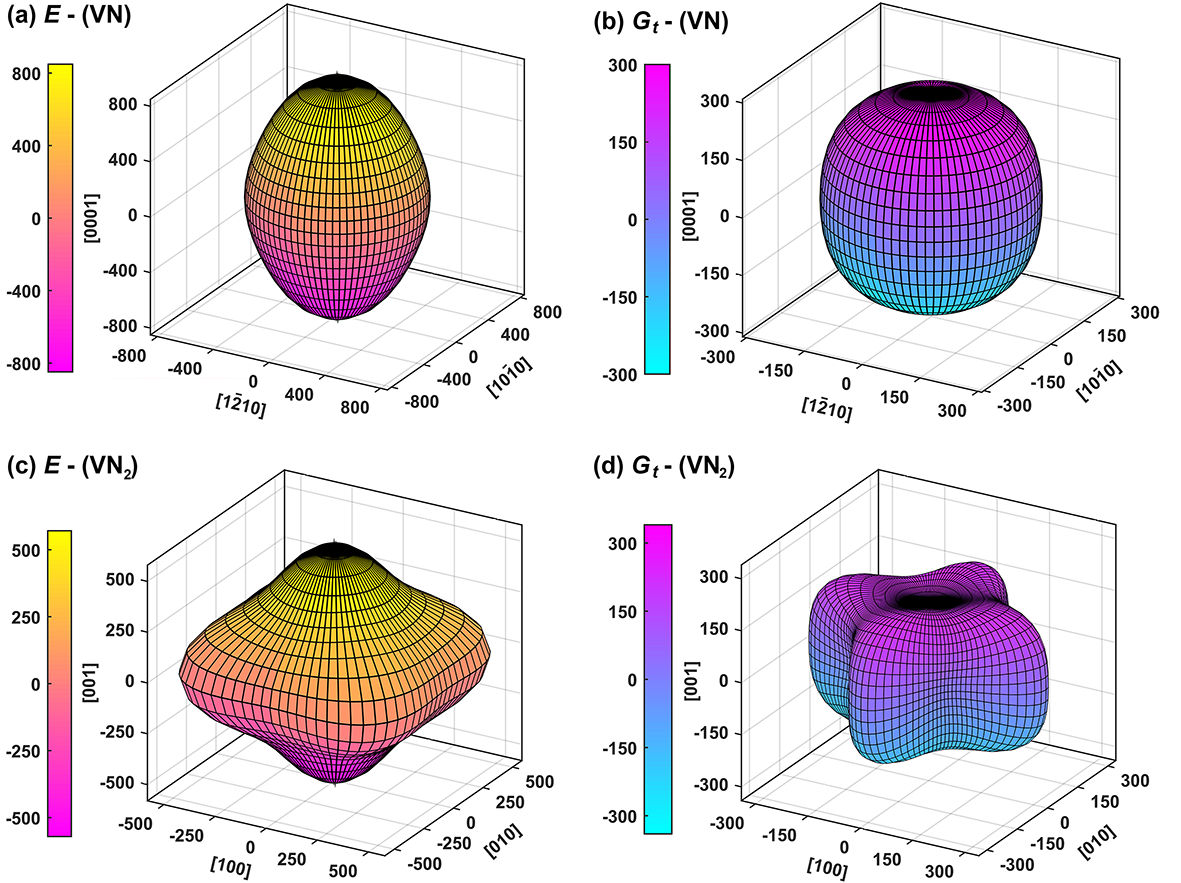}
   \caption{\label{Fig7} (Color online) Illustration of direction dependent of Young's modulus \emph{E} and torsion modulus $\emph{G}_t$ for \emph{P}$\bar{6}$\emph{m}2-VN and \emph{I}4/\emph{mcm}-VN$_2$.}
\end{figure}
\end{center}

Given that WC-type VN exhibits remarkable elastic moduli, it is natural to consider if these outstanding mechanical properties would be found in other WC-type transition metal (TM) mononitrides. Here, we computed elatic moduli, hardnesses and fracture toughness of nitrogen combined with transitional elements from IVB, VB and VIB groups [see Table \ref{Tab3}]. The result indicates that group VB mononitrides exhibit higher hardness and fracture toughness. Bulk moduli increase from IVB to VB to VIB mononitrides, whereas shear moudli peak at VB mononitrides.

\begin{center}
\begin{figure*}
   \includegraphics[angle=0,width=1.0\linewidth]{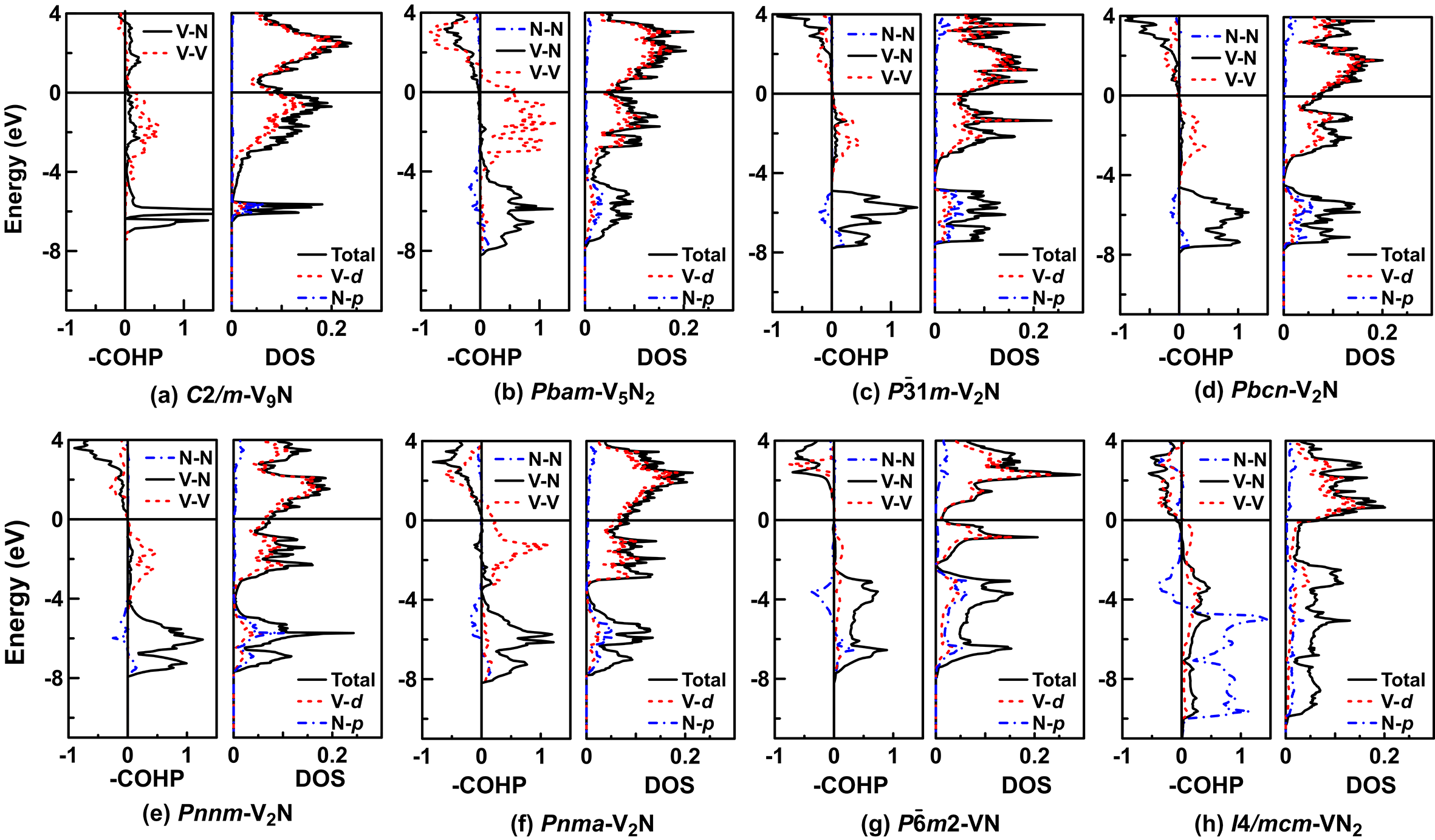}
   \caption{\label{Fig8} (Color online) Total DOS, projected DOS and crystal orbital Hamilton population (-COHP) curves for V-N compounds at 0 GPa. Bonding and antibonding correspond to positive and negative values of -COHP, respectively.}
\end{figure*}
\end{center}

All the crystals are elastically anisotropic, which can be favourable for the emergence of microcracks\cite{ravindran1998density}. Calculating and visualizing the elastic anisotropy is important for understanding these properties and optimizing them for practical applications. The directional dependence of Young's modulus (\emph{E}) and torsion shear modulus ($\emph{G}_t$) for the two hardest V-N compounds (WC-type VN and \emph{I}4/\emph{mcm}-VN$_2$) is displayed in Fig. \ref{Fig7}. A surface construction can be used to clearly reflect the variation of Young's modulus and torsion shear modulus, while it is impossible to represent the direction-dependent shear modulus by three-dimensional diagrams since it is not only related to the direction of the shear plane but also related to the direction of the force. Therefore in engineering, the torsion shear modulus ($\emph{G}_t$, an average in the shear plane), which can be represented by three-dimensional diagrams, is used for visualizing the anisotropy of a material. For an isotropic system, one would see a spherical shape. The degree of elastic anisotropy in a system can be directly extracted from asphericity of these figures. The magnitudes of anisotropy of Young's modulus (\emph{E}) and torsion shear modulus ($\emph{G}_t$) for \emph{I}4/\emph{mcm}-VN$_2$ are greater than that of WC-type VN.

\subsection{Electronic structure and chemical bonding of V-N compounds}
Total and projected density of states (DOS) of V-N compounds are presented in Fig. \ref{Fig8}. All the V-N compounds exhibit metallic behavior and a hybridization between the V-\emph{d} and N-\emph{p} states. Crystal orbital Hamilton population (COHP) is often used to identify the bonding or anti-bounding character of interatomic interactions in a crystal. The calculated -COHP [Fig. \ref{Fig8}] of vanadium subnitrides is characterized by massively V-V and V-N bonding states and minor anti-bonding states between N-N below the Fermi level. It should be pointed out that there is no N-N interaction in \emph{C}2/\emph{m}-V$_9$N due to the absence of direct atomic contacts between N atoms. With the increase of the nitrogen content, the bonding N–N interactions in \emph{I}4/\emph{mcm}-VN$_2$ become strong (as revealed by the positive –ICOHP values) although with small N-N antibonding regions around -3.5 eV [Fig. \ref{Fig8}(h)], which, together with the V-V and V-N bonding states below the Fermi level, greatly enhance the mechanical properties. Compared with other V-N compounds, the large overlap between V-\emph{d} and N-\emph{p} is shown in the projected DOS of WC-type VN. Meanwhile, the -COHP reveals a strong V-N covalent bonding between the V and N (-ICOHP is 2.7 eV/pair for the nearest V-N) and in addition some V-V bonding component [Fig. \ref{Fig8}(g)]. Both evidences correspond to a strong V-N bond in WC-type VN, contributing to its good mechanical properties.

\section{Conclusions}
Stable V-N compounds and their crystal structures at pressures up to 120 GPa have been predicated using \emph{ab initio} evolutionary algorithm USPEX. Four new phases \emph{C}2/\emph{m}-V$_9$N, \emph{Pbam}-V$_5$N$_2$, \emph{Pnma}-V$_2$N and \emph{I}4/\emph{mcm}-VN$_2$ are reported. All the predicted high-pressure vanadium nitrides can be theoretically preserved as metastable phases at zero pressure since they are all dynamically stable at zero pressure. The sequence of phases of V$_2$N under pressure is $\varepsilon$-Fe$_2$N-type V$_2$N $\rightarrow$ $\zeta$-Fe$_2$N-type V$_2$N $\rightarrow$ Fe$_2$C-type V$_2$N $\rightarrow$ \emph{Pnma}-V$_2$N. In addition, the enthalpy and lattice dynamics properties of several vanadium mononitride are analyzed. It is expected WC-type VN can be synthesized under high pressure. Structural features, relative stabilities, mechanical properties, electronic structures and chemical bonding of all the V-N compounds are systematically analyzed at 0 GPa. WC-type VN has the superior Vickers hardness ($\sim$37 GPa) and fracture toughness (4.3-6.1 MPa m$^{1/2}$) which mainly originate from its strong V-N bonding. Another new high-pressure phase \emph{I}4/\emph{mcm}-VN$_2$ also exhibits high bulk modulus (311 GPa), high shear modulus (214 GPa), Vickers hardness (24.8-26.9 GPa) and fracture toughness (3.6-6.6 MPa m$^{1/2}$). Hardness and fracture toughness are determined both by the strength and topology of bond networks.

\section{acknowledgments}
Calculations were carried out the Extreme Science and Engineering Discovery Environment (XSEDE), which is supported by National Science Foundation grant number ACI-1053575. A.R.O. thanks Russian Science Foundation (grant 19-72-30043).

\end{document}